\documentclass[prb,showpacs,twocolumn]{revtex4}
\usepackage{graphicx}
\usepackage{color}

\begin{document}

\title{The superconducting gaps in FeSe studied by soft point-contact Andreev reflection spectroscopy }
\vspace{1cm}

\author{Yu.G. Naidyuk $^{1}$}
\author{O.E. Kvitnitskaya $^{1}$}
\author{N.V. Gamayunova, $^{1}$}
\author{ D.L. Bashlakov $^{1}$}
\author{ L.V.\,Tyutrina $^{1}$}
\author{G. Fuchs $^{2}$}
\author{R. H\"uhne $^{2}$}
\author{D.A. Chareev $^{3,4,5}$}
\author{A.N. Vasiliev $^{5,6,7,8}$}
\vspace{1cm}

\affiliation{$^{1}$ B. Verkin Institute for Low Temperature Physics and Engineering,
National Academy of Sciences of Ukraine, 47 Nauky Ave., 61103, Kharkiv,
Ukraine}
\affiliation{$^{2}$ Institute for Metallic Materials, IFW Dresden, D-01171 Dresden, Germany}

\affiliation{$^{3}$ Institute of Experimental Mineralogy, RAS, 142432 Chernogolovka, Russia}

\affiliation{$^{4}$ Kazan Federal University, 420008 Kazan, Russia}

\affiliation{$^{5}$ Ural Federal University, 620002 Ekaterinburg, Russia}

\affiliation{$^{6}$ Lomonosov Moscow State University, 119991 Moscow, Russia}

\affiliation{$^{7}$ National University of Science and Technology "MISiS", Moscow 119049, Russia}

\affiliation{$^{8}$ National Research South Ural State University, Chelyabinsk 454080, Russia}

\begin{abstract}
FeSe single crystals have been studied by soft point-contact Andreev-reflection spectroscopy.
Superconducting gap features in the differential resistance $dV/dI(V)$ of point contacts such as a
characteristic Andreev-reflection double-minimum structure have been measured versus temperature
and magnetic field. Analyzing $dV/dI$ within the extended two-gap Blonder-Tinkham-Klapwijk model
allows to extract both the temperature and magnetic field dependence of the superconducting gaps.
The temperature dependence of both gaps is close to the standard BCS behavior. Remarkably,
the magnitude of the double-minimum structure gradually vanishes in magnetic field,
while the minima position only slightly shifts with field indicating a weak decrease of the
superconducting gaps. Analyzing the $dV/dI(V)$ spectra for 25 point contacts results in the averaged
gap values $\langle\Delta_{L}\rangle$ = 1.8$\pm $0.4\,meV and  $\langle\Delta_{S}\rangle$=1.0$\pm 0.2$
 meV and reduced values 2$\langle\Delta_{L}\rangle/k_BT_{c}$=4.2$\pm$ 0.9 and 2$\langle\Delta_{S}\rangle/k_BT_{c}$=2.3$\pm 0.5$
for the large (L) and small (S) gap, respectively. Additionally, the small
gap contribution was found to be within tens of percent decreasing with both temperature and
magnetic field. No signatures in the $dV/dI$ spectra were observed testifying a gapless superconductivity
or presence of even smaller gaps.
\end{abstract}



\maketitle

\section{Introduction}
The binary compound FeSe, belonging to the family of iron based
superconductors, is in the focus of intense investigations nowadays. The
main advantage of this material is that superconducting (SC) FeSe is the
only binary compound among this family. Additionally, FeSe shows no long
range magnetic order, which might simplify the understanding of the nature
of SC pairing. Furthermore, FeSe demonstrates extraordinary sensitivity of
the SC properties to external pressure, chemical doping on the Fe or Se site
and to the intercalation by alkaline metals (see [1] for recent reviews).
Besides, the critical temperature of FeSe can be enhanced by an order of
magnitude by diminishing its dimensionality to a 2-D type monolayer.
Further, immensely small Fermi surfaces, which are comparable with the SC
gap(s) $\Delta $, locates SC state of FeSe in the vicinity of the
extraordinary BCS-BEC crossover [2]. Therefore, investigations of SC gap(s)
in FeSe are of high interest.

ARPES is the most powerful method to study the directional and band
dependence of the SC gap(s). However, the resolution of ARPES measurements,
which is nowadays slightly below 1 meV, does not provide sufficient accuracy
for detection of the SC gap value for superconductors with a critical
temperature of about 10 K and below, as in the case of bulk FeSe. Two other
spectroscopic methods as scanning tunneling spectroscopy (STS) [3] and
point-contact Andreev reflection (PCAR) spectroscopy [4, 5] have
significantly better energy resolution, though both methods suffer from the
directional selectivity and especially from the ability to resolve electron
bands.

In one of the first STS measurements on FeSe, Song et al. [6] reported the
presence of one gap with $\Delta \sim $2.2\,meV taken as half of the
peak-to-peak energy separation in tunnel \textit{dI/dV} spectra. Further, Kasahara et al.
[2] demonstrated tunnel \textit{dI/dV} spectra of FeSe showing a V-shaped zero-bias
minimum with the side maxima at $\pm$2.5\,meV and shoulders at $\pm$3.5\,meV.
These features were taken as an evidence of two SC gaps. Later, Watashige et
al. [7] reported STS data and \textit{dI/dV} spectra with peaks at $\pm$2.5\,meV and
shoulders outside of the main peaks at $\pm$3.5\,meV. Moore et al. [8] obtained
a V-shaped STS spectrum for FeSe in the low energy range with clear peaks at
$\Delta =\pm $2.3 meV. Recently, Jiao et al. [9] used the ($s $+\textit{es}) model
to fit their STS data with a small $s$-wave gap of $\Delta _{s}$(0)=0.25 meV
and a large anisotropic extended $s$-wave gap $\Delta _{es}=\Delta
_{0}$(1+$\alpha \cos 4\Theta )$ with $\Delta _{0}$=1.67 meV and $\alpha $=0.34, what
results in the SC gap maximum of 2.24 meV and a minimum of 1.10 meV.

At the same time, one of the last ARPES studies of FeSe reported by
Borisenko et al. [10] has announced two gaps equal to 1.5 and 1.2 meV for
the hole band in the center and for the electron band in the corner of the
Brillouin zone, respectively. Here, we must note that recently Hong {\&}
Aberge [11] pointed out that the side peaks observed in STS measurements on
compounds with strong electron-electron correlation, like iron-based
superconductors and high-Tc superconductors, are formed by
coherence-mediated tunneling under bias. Because of that, such peaks do not
reflect directly the underlying density of states (DOS) of the sample and
the gap measured between side peaks observed in STS is bigger than the SC
gap observed by ARPES. This might be the reason of substantial differences
in the mentioned gap values obtained by STS [2,6,7,8] and ARPES [10].

Turning to the Andreev-reflection spectroscopy of SC gaps in FeSe, Ponomarev
et al. [12] have detected two sets of subharmonic gap structures due to
multiple Andreev-reflection using break-junctions with polycrystalline
samples. This was taken as proof of two nodeless SC gaps $\Delta
_{L}$=2.75$\pm $0.3 meV and $\Delta _{S}$=0.8$\pm $0.2 meV. At the same
time, their result on the temperature dependence of the both gaps was
curious. Later, they reported new values $\Delta _{L}$=2.4$\pm $0.2 meV
and $\Delta _{S}$=0.75$\pm $0.1 meV using single crystals [13]. In our
PCAR measurements with FeSe single crystals [14], we also extracted two gaps
from measured \textit{dV/dI} with gap values similar to those in STS experiments, though
the contribution of the larger gap at 3.5 meV to the total PC conductivity
was rather small, of order of 10{\%}.

As follows from all the above points, there is a challenge to determine the
spectral data related to the value of the SC gaps more accurately. Besides
it, there is lack of data in the literature for the temperature and
especially magnetic field dependence of the SC gap(s) in FeSe. All these
issues are target of the current investigation of FeSe using the technique
of PCAR spectroscopy.

\section{Experimental details}
The plate-like single crystals of FeSe$_{1-x}$ ($x$=0.04 $\pm$0.02) were grown
in evacuated quartz ampoules using a flux technique as described in [13].
The resistivity and magnetization measurements revealed a SC transition
temperature up to $T_{c}$=9.4\,K. So called ``soft'' method was utilized to
create point contacts, i.\,e. a tiny drop of silver paint was placed on the
freshly cleaved surfaces of FeSe. The soft PC's were made on the \textit{ab}-plane
cleaved with the scalpel or on the edge of a thin FeSe flake. We will refer
to these two types of PC's as a ``plane'' or ``edge'' PC, respectively. The
silver paint drop was connected to the electrical circuit by Cu, Ag or Pt
thin wires with a diameter of 0.1 mm or slightly less. The size of the
silver paint drop was about several hundred microns, while the PC resistance
between the silver paint drop and FeSe samples was usually in the range 0.5--10\,$\Omega$.
Such resistance corresponds to the PC size of the order of several tens of
nanometers [4] in the case of PC between ordinary metals. Therefore, it is
assumed for our case, that either there is a large number of nanometer-sized
PC's, or the interface between silver paint and FeSe has some barrier (e.\,g.,
oxide). In spite of the unknown microscopic picture of the real PC
structure, the actual shape of \textit{dV/dI} characteristics is more important. As we
will demonstrate below, \textit{dV/dI} show typical Andreev-reflection SC gap related
features, which we call double-minimum structure.

The differential resistance $dV/dI(V) \equiv  R(V)$ of the PC was recorded by sweeping
the $dc$ current $I$ on which a small $ac$ current $i$ was superimposed using a standard
lock-in technique. The measurements were performed in the temperature range
from about 3\,K to above $T_{c}$ and in magnetic field up to 15\,T, applied both
along to the \textit{ab}-plane or parallel to the $c$-axis.

\begin{figure}[t]
\includegraphics[width=0.45\textwidth]{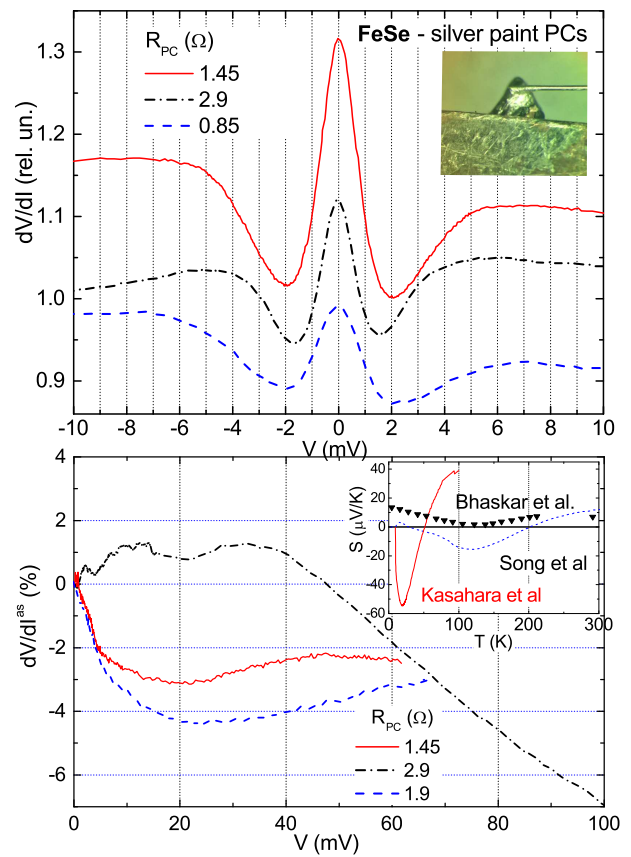}
\caption{(Color online) Upper panel: Examples of raw \textit{dV/dI} curves measured at 3 K for three
``soft'' PC's created by a tiny drop of silver paint on a cleaved FeSe
surface. The small picture shows an image of the FeSe single crystal
(triangle shape) with a drop of silver paint and a Cu wire with a diameter
of 0.08 mm. Bottom panel: antisymmetric part \textit{dV/dI}$^{as}$({\%})
=100[\textit{dV/dI(V$>$}0) - \textit{dV/dI}(V$<$0)]$/$2\textit{dV/dI}($V$=0) of \textit{dV/dI} calculated for the PC's from the upper panel.
Inset shows the behavior of thermopower in single FeSe crystals according to
Kasahara et al. [2] and FeSe polycrystals reported by Song et al. [16] and
Bhaskar et al. [17].
\label{fig:fig1}}
\end{figure}

\section{Results}
Fig.\,1a shows the \textit{dV/dI }spectra for several soft PCs, which demonstrate a
characteristic double-minimum structure with the minima position between
1.5 -- 2\,mV, which is close to the expected SC gap value. We note the perfect
reproducibility of the SC features in \textit{dV/dI}, i.\,e. almost all of the more than
twenty soft PC's with resistance in the range 0.4 -- 5$\Omega $ show the
pronounced double-minimum structure in\textit{ dV/dI}. It is in contrast to our previous
measurements on the same FeSe crystals using a needle-anvil geometry with
tips from Cu, Ag or W thin wires [14, 15], where the double-minimum
structure in \textit{dV/dI} appeared very rarely. At the same time, soft PC's with the
higher resistance display \textit{dV/dI} with weak zero bias minimum or
absence of any SC features at all, similar to the needle-anvil type PC's
shown on Fig.\,1 in [14].

The \textit{dV/dI }curves in Fig.\,1a are asymmetric with an enhanced value at negative bias
similar to our previous data in [14], so that the calculated antisymmetric
part of \textit{dV/dI} (\textit{dV/dI}$_{ }^{as}$({\%})=100[\textit{dV/dI(V$>$}0) - \textit{dV/dI}(V$<$0)]$/$2\textit{dV/dI}($V$=0) is negative (see
Fig.\,1b). At the same time, about one third of the PC's demonstrates positive
\textit{dV/dI}$_{ }^{as}$ at low bias as shown in Fig.\,1b.

\begin{figure}[tbp]
\includegraphics[width=0.5\textwidth]{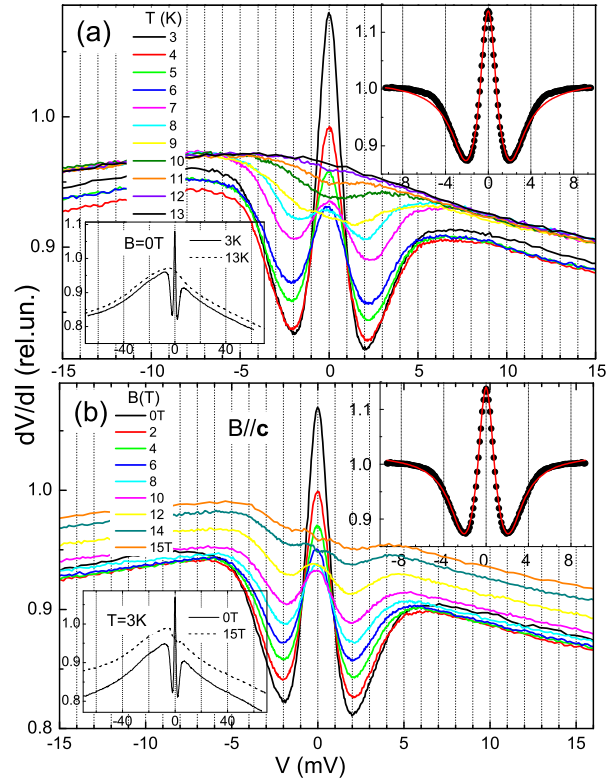}
\caption{(Color online) Temperature (panel a) and magnetic field (panel b) evolution of the
\textit{dV/dI} spectra of a soft FeSe PC with normal state resistance of 1.45\,$\Omega $.
Left insets on both panels show \textit{dV/dI }at larger bias taken at low temperature and
above the critical temperature (panel a) or at maximal magnetic field (panel
b). Right inset in panel (a) shows the fit of the symmetrized \textit{dV/dI} at 3\,K
normalized to the normal state (see text and Table 1, PC{\#}552) by the
two-gap Blonder--Tinkham--Klapwijk model, while right inset in panel (b)
shows the fit of the same \textit{dV/dI} by an anisotropic model $\Delta=\Delta_0
(1+\alpha \cos 4\Theta )$ (see text).
\label{fig:fig2}}
\end{figure}

Fig.\,2 demonstrates the evolution of \textit{dV/dI} curves for a soft PC versus
temperature and magnetic field. We used these data to determine the SC gap
value and its temperature and magnetic field dependence. Therefore, we
fitted \footnote{As mentioned in the introduction, the Fermi energy of FeSe
is comparable to the value of the SC gap(s). This puts into the question, whether
the BTK or similar existing models can be applied in order to extract the
SC gap. However, due to lack of alternative, we applied the BTK model. This
model fits \textit{dV/dI} almost perfectly (see inset in Fig. 2a). Anyway, such situation
must be analyzed theoretically to be sure that, at least, the BTK model can
be used, even in the case of E$_{F} \sim \Delta$. It should be noted
that the BTK equations for the current trough a PC contain BCS quasiparticle
DOS. As it is shown in [25], the DOS calculated at T = 0\,K for the model
with one hole and one electron pockets (as in FeSe) has similarity with the
BCS DOS, but contains additional singularities (steps) for the hole band
(see Fig.\,11 in [25]). However, these singularities will be smeared out at a
temperature increase, so expectedly the DOS will be more similar to BCS
shape for increasing temperature. Probably, this is the reason of the
overall quite good fit using the BTK equations with standard BCS DOS.}the
\textit{dV/dI} curves normalized to the normal state using the two-gap
Blonder--Tinkham--Klapwijk (BTK) model (see, e.g. [4, 5] for some details of
the fit for different models and superconductors). An example of the fit for
the \textit{dV/dI} at 3\,K is shown in the inset of Fig.\,2a. The fit is perfect,
excluding small deviations between 4 and 8\,meV, where so-called humps or
side-maxima occur, which arise from a non-Andreev-reflection contribution to
the \textit{dV/dI} spectra. The results of the SC gap behavior after the fit procedure are
presented in Fig.\,3. The gap values at 3\,K are $\Delta _{L}\approx $1.9
and $\Delta _{S} \approx $1.0\,meV for the large (L) and small (S) gap,
respectively, with an about 80{\%} contribution to the \textit{dV/dI} coming from the
large gap\footnote{ According to the low-temperature specific heat
measurements in [26], contribution of the large gap is estimated to 71{\%}
within two-gap model.}. The extracted gap values correspond to a 2$\Delta
/k_{B}T_{c}$ ratio of about 4.2 and 2.2 for the large and the small gap,
respectively, if we use a $T_{c}$ = 10.5\,K obtained from the BCS
extrapolation in Fig.\,3a. The temperature behavior of both gaps is close to
the BCS-like curve, while the contribution of the small gap to \textit{dV/dI} spectra
decreases with increasing temperature. The contribution of the small gap to
\textit{dV/dI} also vanishes in magnetic field, while both gap values are only weakly field
dependent. It is difficult to specify the critical temperature or magnetic
field, at which the small gap contribution disappears due to diminution and
smearing of all ``gap'' structures with increasing temperature or magnetic
field. This makes the fit procedure less unambiguous. The fit parameters for
\textit{dV/dI} of several PC's measured at 3 K are shown in Table 1.

\begin{figure}[tbp]
\includegraphics[width=0.5\textwidth]{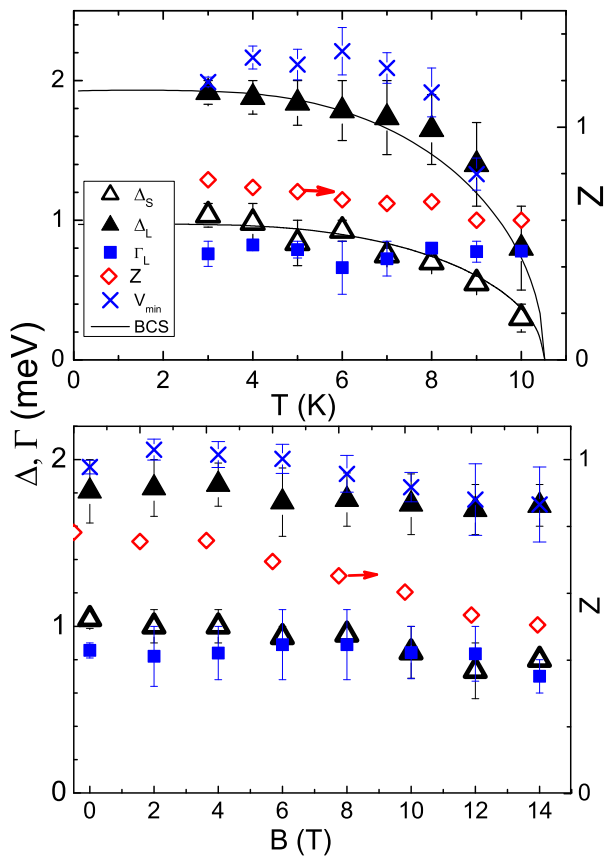}
\caption{Temperature and magnetic field dependencies of the fitting parameters for the
PC from Fig.\,2 in a two-gap approximation: closed and open triangles are the
large and small gaps $\Delta _{L }$ and $\Delta _{S}$, respectively,
$\Gamma_{L}$ is the broadening parameter (squares), Z is the barrier parameter
(diamonds), V$_{min}$ is the minimum position in \textit{dV/dI }(crosses).
\label{fig:fig3}}
\end{figure}

\begin{table*}[tbp]
\begin{ruledtabular}
\begin{center}
\caption{Fit parameters for \textit{dV/dI} of several soft PC's measured at 3\,K:
R$_{PC}$ is the PC resistance, V$_{min}$ is the minimum position in
\textit{dV/dI}, $\Delta _{L,S }$ is the large and small SC gaps, $\Gamma _{L}$ is the
broadening parameter, Z is the ``barrier'' parameter, $ w$ is the weight factor
(contribution to \textit{dV/dI}) of the small gap, $\Delta _{aver}$= (1-$w)\Delta
_{L}+w\Delta _{S}$,$_{ }$ S is the scaling parameter. $\Gamma _{S}$
for the small gap is taken to be zero. Bold name marks those PC's which
\textit{dV/dI} are shown in Fig.\,1.}
\vspace{0.5cm}
\begin{tabular}{lllllllllll}
Name&R$_{PC} \par (\Omega) $&Type&V$_{min}$ \par (mV)&$\Delta _{L}$ \par (meV)&$\Delta _{S}$ \par (meV)&$\Delta _{aver}$ \par (meV)&Z&$\Gamma_{L}$ \par(meV)&$w$&S \\
\hline
\textbf{{\#}351}&0.85&plane&1.5&1.5&0.7&1.33&0.70&0.5&0.22&0.27 \\
{\#}373&0.67&plane&1.75&1.73&0.9&1.51&0.77&0.55&0.26&0.19 \\
{\#}503&1.2&plane&1.6&1.46&0.8&1.4&0.68&0.77&0.09&1.5 \\
{\#}608&3&plane&1.7&1.53&0.8&1.49&0.68&0.87&0.06&1.18 \\
{\#}401&1.0&edge&1.5&1.5&0.7 &1.32&0.72&0.74&0.22&0.36 \\
\textbf{{\#}416 }&2.9&edge&1.6&1.62&0.84&1.43&0.74&0.35&0.28&0.43 \\
\textbf{{\#}552}&1.45&edge&2&1.9&1.0&1.74&0.77&0.76&0.18&0.9 \\
\end{tabular}
\label{tab1}
\end{center}
\end{ruledtabular}
\end{table*}

\begin{center}
\textbf{Discussion}
\end{center}

We formed soft PC's in two ways: (a) placing a silver paint drop on the
plane (flat surface) of the FeSe flake or (b) on the edge of the thin flake.
In the last case the silver paint also partially covered the flat surface,
because the drop size was larger than the flake thickness. Probably due to
this effect, we did not observe a big difference in the shape of \textit{dV/dI} and there
was also no notable difference between the gap values for the ``edge'' and
``plane'' PCs.

Here, it is appropriate to imagine the picture how tiny PC's will be formed by
dripping silver paint onto the FeSe surface. Let`s take into consideration
that according to [18], FeSe single crystals have a huge anisotropy in
resistivity between the $c$-axis and \textit{ab}-plane, typically $\rho _{c}$/$\rho
_{ab} \quad \approx $ 500 below the SC transition. In such case, we assume
that the conductivity between the silver drop and the flat FeSe surface,
that is along the $c$-axis, is minor and the current flows mainly through the
edge of terraces on the surface covered by the silver paint, which opens
channel(s) to the $ab$-plane. Thus, despite the large
contact area covered by the silver paint, the current flows mainly through
the confined area at the edge of the terraces. As such, it does not matter
whether we prepared PC on plane or edge of FeSe flake, in both cases the
current preferably flows within the $ab$-plane and no remarkable anisotropy is
expected.

Let us discuss details of the fit procedure. The two-gap fit uses, in
general, 7 parameters. Among them are two gaps $\Delta _{L,S}$, two
broadening parameters $\Gamma_{L,S}$, two barriers Z$_{L,S}$ and the weight
factor $w$. In the case, when the \textit{dV/dI} spectrum shows only a single double-minimum,
it leaves a wide scope or ``too much room'' for the fitting parameters and
makes the fit controversial. Therefore, we shortened the number of the
fitting parameters supposing equal barriers for the both gaps Z$_{L
}$=Z$_{S}$. Additionally, we supposed $\Gamma_{S}$=0 taking into account the
minor small gap contribution. Thus, the number of fit parameter was reduced to
5. Obviously, some variation of the extracted data is still possible even
using five fitting parameters, however the gap(s) value(s) must concentrate
around the minima position of about 1.5--2\,meV in any case. This is seen
also from the columns {\#}4 -- 7 in the Table\,1. The average values for
the gaps for several PC's presented in Table\,1 are $\Delta _{L}$=1.6 and
$\Delta _{S}$=0.8 meV, so that the large gap value is close to that
measured by ARPES [10], while the small gap value is about 30{\%} smaller
than the ARPES data. On the other hand our gap values are smaller than the
gap maximum 2.24 meV and gap minimum 1.1 meV reported in [9]. However, the
gap ratio $\Delta _{L}$/$\Delta _{S}$ is close to 2 in both cases. At
the same time, we did not observe neither gap-features in \textit{dV/dI} similar to
$\Delta _{S}$=0.25 meV, as reported in [9], nor gapless \textit{dV/dI} behavior (like
single V-shaped zero-bias \textit{dV/dI} minimum). All of \textit{dV/dI} data from our soft PC's demonstrate
zero-bias maximum, as shown, e.\,g., in Fig.\,1a.

The temperature dependence of the large gap is close to the BCS behavior.
The contribution $w$ of the small gap decreases with increasing temperature.
The larger gap value is only weakly field dependent. The latter is in line
with the observed minima positions in \textit{dV/dI}, which are only slightly reduced in
magnetic field, despite of the overall vanishing of the double-minimum
structure. Also, the contribution $w$ from the small gap decreases with
increasing magnetic field. We observed a similar weak magnetic field
dependence of the SC gap for another multiband superconductor from the
nickel-borocarbide family, namely, TmNi$_{2}$B$_{2}$C [19]. There, the
possible interpretation of the observed gap behavior versus magnetic field
was related to a multiband scenario. Additionally, the electronic DOS
modification in the mixed state and vortex pinning near the contact
interface were suggested. However, such magnetic field gap behavior is still
not completely understood.

\begin{figure}[t]
\includegraphics[width=0.5\textwidth]{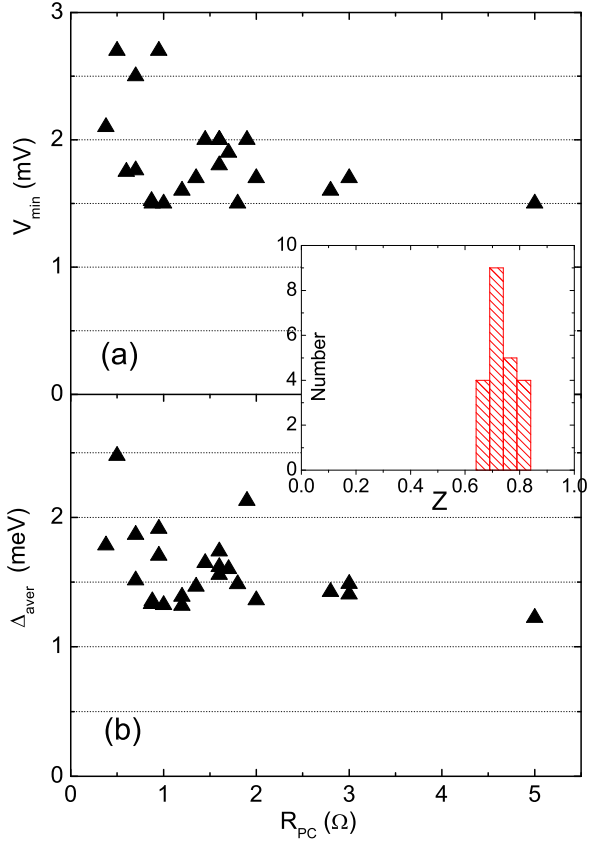}
\caption{(a) Distribution of the minima position in \textit{dV/dI} with double-minimum
structure for all soft PCs. (b) Distribution of the averaged gap calculated
for each PC according to equation $\Delta _{aver}$=(1-$w)\Delta
_{L}+w\Delta _{S}.$ Inset shows a histogram for the distribution of Z for
all PCs.
\label{fig:fig4}}
\end{figure}

We fitted our data also by the anisotropic gap model with $\Delta =\Delta_0
(1+\alpha  \cos 4\Theta )$ [9]. Additionally, we included in this model a smearing
parameter $\Gamma = \Gamma_{0 }(1+\alpha \cos 4\Theta )$ as well. The description of the
experimental data by this model is also fine (see Fig.\,2b inset). The
fit results in $\Delta $=1.42\,meV, $\alpha $=0.6, Z=0.81 and $\Gamma$=0.53\,meV. The
extracted temperature dependence of $\Delta $(T) almost perfectly follows
the BCS dependence. However, the monotonic increase of $\alpha $ with
temperature up to the maximal value 1 close to T$_{c}$ is not physically
reasonable in this case, whereas on the other hand $\Gamma_{0 }$ goes down to
zero. If we try to keep \textit{$\alpha $} more or less constant, the fit becomes worse and
$\Delta _{0 }$ slowly increases with temperature before to drop at approaching $T_c$.
So, our conclusion is that the anisotropic $\alpha $-model is less compatible with our data.

Let us look close on the statistics of the data after analyzing 25 soft
PC's. Fig.\,4a shows the distribution of the minimum position $V_{min}$ in
\textit{dV/dI} for all PC's. The $V_{min}$ data agglomerate in the range between
1.5--2\,mV with an average value of $\langle V_{min}\rangle$=1.75$\pm $0.25\,mV. Several
PC's with low resistance exhibit a larger $V_{min }$, what might be due to
the influence of some small serial or spreading resistance. Analyzing of
\textit{dV/dI} PCAR spectra for all PC's results (see Table\,2) in gap values of
$\langle\Delta_{L}\rangle$=1.8$\pm $0.4\,meV and $\langle\Delta_{S}\rangle$1.0$\pm $0.2\,meV for
the large (L) and small (S) gap, respectively, leading to reduced gap values
of 2$\langle\Delta_{L}\rangle/k_{B}$T$_{c}$=4.2$\pm $0.9 and
2$\langle\Delta_{S}\rangle/k_{B}$T$_{c}$=2.3$\pm $0.5. Here, we used an averaged
$T_{c}\approx $10\,K value obtained by fitting of the temperature
dependence of the gap by a BCS-like curve. Fig.\,4b shows the calculated mean
gap value $\Delta _{aver}$=(1-$w)\Delta _{L}+w\Delta _{S}$ resulting
from the fit of \textit{dV/dI} curves at 3\,K for all PCs. Here, $\Delta _{aver}$ is
between 1.3 and 1.9\,meV excluding 3 marginal PC's with higher $V_{min}$. In
this case, the averaged value for all PC's is $\langle\Delta _{aver}\rangle$ =
1.6$\pm $0.3\,meV. As a result, we received an averaged ratio 2$\langle\Delta
_{ aver}\rangle/k_{B}T_{c}$=3.7$\pm $0.7, which is a bit higher than the
BCS value 3.52.

\begin{table*}[htbp]
\begin{ruledtabular}
\begin{center}
\caption {Averaged data after analyzing \textit{dV/dI} at 3\,K for 25 soft PC's
within the BTK model:
$V_{min}$ is the minimum position in \textit{dV/dI}, $\langle\Delta _{L,S}\rangle$ is the
average of large and small gaps, Z is the ``barrier'' parameter,  $\Delta
_{aver}$= (1- $w)\Delta _{L}+w\Delta _{S}$. $T_{c}$ is taken equal to
10\,K. Note, the data for 8 soft PC's created on new, so
called, 3D FeSe samples are also included in statistics. The latter look like bulky pieces, contrary to
plate shaped usual FeSe flakes.}
\vspace{0.5cm}
\begin{tabular}{p{54pt} p{43pt} p{43pt} l l p{44pt} l l l }
$\langle V_{min}\rangle$ \par meV&$\langle\Delta_{L}\rangle$ \par meV&$\langle\Delta_{S}\rangle$ \par meV&Z&$w$&$\langle\Delta_{aver}\rangle$ \par meV&
2$\langle\Delta_{L}\rangle$/k$_{B}$T$_c$&2$\langle\Delta_{S}\rangle$/k$_B$T$_c$&2$\langle\Delta_{aver}\rangle$/k$_B$T$_c$ \\
\hline
1.75$\pm $0.25&1.8$\pm $0.4&1.0$\pm $0.2&0.7$\pm $0.1&0.17$\pm $0.13&1.6$\pm $0.3&4.2$\pm $0.9&2.3$\pm $0.5&3.7$\pm $0.7 \\
\end{tabular}
\label{tab2}
\end{center}
\end{ruledtabular}
\end{table*}

According to the latest data from [20], where the authors used sub-kelvin
Bogoliubov quasiparticle interference (BQPI) imaging, ``the maximum gaps
were assigned to each band based on the energy evolution of BQPI to the
energy limit E$\to $2.3meV for the $\alpha $-band and E$\to $1.5meV for the
$\varepsilon $-band''. These values are larger comparing to our data for the
large and small gaps. At the same time, the authors of [20] found an extraordinarily
anisotropic ($\Delta _{\alpha }^{max}$/$\Delta _{\alpha
}^{min}\mathbin{\lower.3ex\hbox{$\buildrel>\over
{\smash{\scriptstyle\sim}\vphantom{_x}}$}} $15) C$_{2}$-symmetric energy-gap
structure. Apparently, our data for the large and small gaps represent the
averaged gap for the corresponding $\alpha $- and $\varepsilon $-bands.
Finally, we show in Table III existing data for the gap(s) in FeSe measured by different spectroscopic
methods.

Take a note on the distribution of Z values for all PC's in the inset of
Fig.\,4. Curiously, the Z values have a low spreading and concentrate around
0.7$\pm $0.1. The low dispersion of Z testifies in favor of some natural
barrier, probably of semiconducting origin\footnote{ As it is discussed in our
previous paper [14], "semiconducting" behavior of \textit{dV/dI} (\textit{dV/dI}
decreases with a bias above the gap structure) can be due to the low
concentration of carriers and/or depleted (semiconducting) surface layer,
violation of stoichiometry and the distribution of Fe vacancies, formation
of oxide on the cleaved surface of FeSe under air exposure. Therefore, we
believe that the ``semiconducting'' \textit{dV/dI} shape is due to oxide or the degraded
surface layer playing a role of weak tunnel barrier. }. Thus one can also consider
the ``semiconducting'' type of the \textit{dV/dI} background. Besides, as it is seen from
Fig.\,2b, Z slightly decreases with temperature, that is expected in the case
of low barrier heights. Therefore, if it is really a natural barrier, then
our assumption Z$_{L}$=Z$_{S}$ in the fit procedure is justified. However,
why Z decreases in a magnetic field (see Fig.\,3) is not yet understood.

Now, we turn to the antisymmetric part of \textit{dV/dI}$^{as}$, which is shown for some
investigated soft PCs in Fig.\,1b. We related the asymmetry of the \textit{dV/dI}
characteristics to thermopower effects in the case of heterocontacts in the
thermal regime [21]. In this case, the antisymmetric part of \textit{dV/dI}$^{as}$ is
proportional to the difference between the Seebeck coefficients $S(T)$ of the
contacting materials [22]. Such correspondence between \textit{dV/dI}$^{as}$ and the
Seebeck coefficient $S(T)$ was observed for PC measurements on [1111] and [122]
iron-based superconductors (see [23, 24]). Additionally, we reported such
correlation also for FeSe in Ref.\,[14]. Our soft PC's mainly had a negative value
of \textit{dV/dI}$^{as }$and only about one third of all PC's exhibited a positive
\textit{dV/dI}$^{as }$prior the \textit{dV/dI}$^{as}$ sign changes (see Fig.1b). Here, we must pay
attention, that the Seebeck coefficient $S(T)$ in FeSe measured by different
authors varies in value, shape and sign (see, e.\,g., inset in Fig.\,1b). In
addition, $S(T)$ of FeSe polycrystals measured in [17] is even positive for
temperatures up to 500 K. We see in the inset of Fig.\,1b a remarkable
difference of $S(T)$ between single crystals and polycrystals as well as also a
different sign of $S(T)$ for two polycrystals. This can be the reason of such
variety of \textit{dV/dI}$^{as}$ for different PCs. Taking into account the huge
anisotropy of resistivity of FeSe according to [18], it is not excluded that
thermopower measured along the $c$-direction can have also different behavior
and sign.

\begin{table}[htbp]
\begin{ruledtabular}
\begin{center}
\caption {Literature data as to the SC gap(s) in FeSe. Jiao et al. [9] and Sprau et al. [20]
reported about anisotropic gap(s).}
\vspace{0.5cm}
\begin{tabular}{lllll}
Method & $\Delta _1$(meV) & $\Delta _2$(meV) & Refs. \\
\hline
STS&2.2&  & Song et al. [6]\\
STS&2.5&3.5  & Kasakhara et al. [1]\\
STS&2.3& &  Moore et al. [8] \\
STS&0.25&1.1-2.24 anis.  & Jiao et al. [9]\\
ARPES&1.2&1.5  & Borisenko et al. [10]\\
MAR&0.8&2.75  & Ponomarev et al. [12]\\
QPI&0.5-1.5 anis.&0.5-2.2 anis.  & Sprau et al. [20]\\
PC&2.5&3.5 &  Naidyuk et al.[14]\\
Soft PC&1$\pm $0.2&1.8$\pm $0.4 &  This work \\
\end{tabular}
\label{tab2}
\end{center}
\end{ruledtabular}
\end{table}

\section{Conclusion}
We investigated SC gaps in FeSe single crystals using soft PCAR
spectroscopy. We measured \textit{dV/dI} with the characteristic for PCAR double-minimum
structure versus temperature and magnetic field for about 25 PCs. Soft PC's
were created by placing a tiny drop of silver paint on the cleaved surface
of FeSe or on the edge of FeSe flake, what assumes the contacts formation
along the c-axis and the base plane. However, there was no noticeable
anisotropy in \textit{dV/dI} spectara observed. Analysis of \textit{dV/dI} data by the extended two-gap
BTK model allows to extract the temperature and magnetic field dependence of
the SC gaps. The temperature dependence of the both gaps is close to the
standard BCS behavior. The PCAR double-minimum structure gradually decreases
in magnetic field. Nevertheless, the position of the minima has a weak field
dependence, leading to almost field independent SC gaps value. This
observation is still not completely understood. Analysis of \textit{dV/dI} PCAR spectra
for all PC's results in gap values of $\langle\Delta_{L}\rangle$=1.8$\pm $0.4\,meV
and $\langle\Delta_{S}\rangle$=1.0$\pm $0.2 meV for the large (L) and small (S)
gap, respectively, what leads to the reduced gap values of 2$\langle\Delta_{L}\rangle/k_{B}T_{c}$=4.2$\pm $0.9
and 2$\langle\Delta_{L}\rangle/k_{B}T_{c}$=2.3$\pm $0.5. At the same time, the small gap
contribution to the spectra is somewhere within 10-20{\%}. Additionally, the
averaged gap value $\Delta _{aver}$=(1-$w)\Delta _{L}+w\Delta _{S}$ for
all PC's amounts to 1.6$\pm $0.3 meV, so that the averaged ratio is
2$\langle\Delta _{ aver}\rangle/k_{B}T_{c}$= 3.7$\pm$0.7, only a bit higher
than the BCS value 3.52. No features in \textit{dV/dI} spectra to testify for the presence
of a gapless superconductivity or the presence the gap smaller than
extracted from the analysis were observed.

\section*{Acknowledgements}
Yu.G.N., O.E.K., N.V.G., D.L.B. acknowledge support of Alexander von
Humboldt Foundation in the frame of a research group linkage program,
partial support of Volkswagen Foundation and funding by the National Academy
of Sciences of Ukraine under project $\Phi $3-19. Yu.G.N., O.E.K., N.V.G., D.L.B.
would like to thank IFW Dresden for hospitality and K. Nenkov for technical
assistance. A.N.V. acknowledges support of the Ministry of Education and
Science of the Russian Federation in the frames of Increase Competitiveness
Program of NUST ``MISiS'' (¹ Ê2-2016-066). D.A.C and A.N.V were supported by
Act 211 Government of the Russian Federation, contracts 02.A03.21.0004,
02.A03.21.0006 and 02.A03.21.0011. G.F. acknowledges support of the German
Federal Ministry of Education and Research within the project ERA.Net RUS
Plus: No146-MAGNES financed by the EU 7$^{th}$ FP, grant no 609556.

\end{document}